\documentclass[prl,superscriptaddress,floatfix,twocolumn,showpacs,amsfonts,amsmath,amssymb,final]{revtex4}
\usepackage{graphicx} 
\usepackage{dcolumn} 
\usepackage{bm} 
\usepackage{color}

\usepackage{tikz}
\usepackage{physics}
\usepackage{siunitx}

\begin{document}

\title{Valley separation via trigonal warping}

\author{Samuel S. R. Bladwell}
\affiliation{School of Physics, University of New South Wales, Sydney 2052, Australia}

\begin{abstract}

Monolayer Graphene contains two inequivalent local minimum, valleys, located at $K$ and $K'$ in the 
Brillouin zone. There has been considerable interest in the use of these two valleys as a doublet 
for information processing. Herein we propose a method to resolve valley currents spatially, using only a 
weak magnetic field. Due to the trigonal warping of the valleys, a spatial offset appears 
in the guiding centre co-ordinate, and is strongly enhanced due to collimation. 
This can be exploited to spatially separate valley states. Based on current experimental devices, spatial 
separation is possible for densities well within current experimental limits. Using numerical simulations, 
we demonstrate the spatial separation of the valley states.

\end{abstract}

\maketitle

{\it Introduction}: Due to the particular symmetry of the honeycomb lattice of 
monolayer graphene, the valence and conduction bands meet at 6 points.
In the immediate 
vicinity of these points, the dispersion is linear and the Fermi surface consists of 
two inequivalent cones at points $K$ and $K'$ in the Brillouin zone. These valleys are 
independent and degenerate, and several works have proposed using these valleys 
as a doublet for information processing; referred to as valleytronics, in analogy with spintronics\cite{Rycerz2007, Schaibley2016}. 
Since the first proposal over a decade ago, considerable progress has been made 
with regards to the generation of both static valley polarisations, and 
valley polarised currents\cite{Garcia2008, Gunlycke2011, Jiang2013, Settnes2016}. Detecting valley polarisation on the other hand has proved to 
be difficult. An early proposal suggested using superconducting superconducting contacts\cite{Akhmerov2007}. 
More recently it has been shown that static valley polarisations can be induced and detected 
via second harmonic generation\cite{Golub2011, Wehling2015}. Nonetheless, the detection of 
valley polarised currents remains an ongoing challenge, with implications for a wide variety of 
phenomena beyond valleytronics. 

In this letter, motivated
by recent developments in electron optics in graphene
we propose an approach for the  detection of valley polarised currents in graphene. 
Over the past 
decade, a variety of improvements in material processing have allowed for high mobilities, with mean free paths
 of tens of microns\cite{Banszerus2016}. 
 Very recently, several groups have considered how to form highly collimated electron 
beams in graphene; Barnard {\it et al} using absorptive metal contacts to form a pinhole aperture, 
and Liu {\it et al} using a parabolic {\it p-n} junction as a refinement of the Veselago lens\cite{Barnard2017, Liu2017}. 
Herein we show that collimation, 
combined with the trigonal warping of the Dirac cone, results in a significant enhancement in the spatial 
separation between ballistic valley polarised currents. 
Combined with an appropriate device layout, this spatial separation can be 
exploited to individually address distinct valley states.
Due to the significant enhancement, we find that the required trigonal warping is small, and the 
required density is well within current experimental limits. 
It thereby provides a novel method of detecting 
ballistic
valley polarised currents.

\begin{figure}[t!]
    \begin{center}
     {\includegraphics[width=0.11\textwidth]{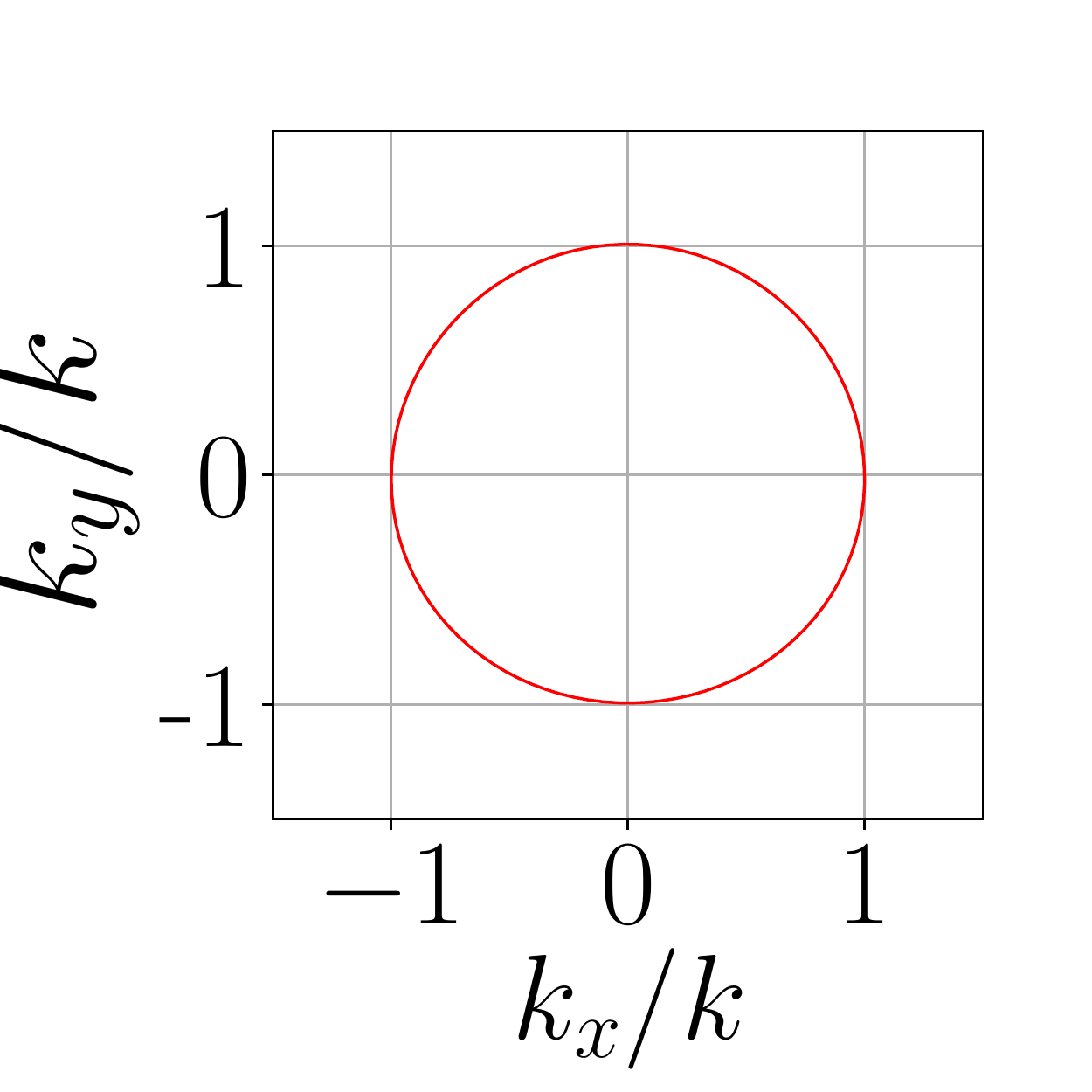}}
     {\includegraphics[width=0.11\textwidth]{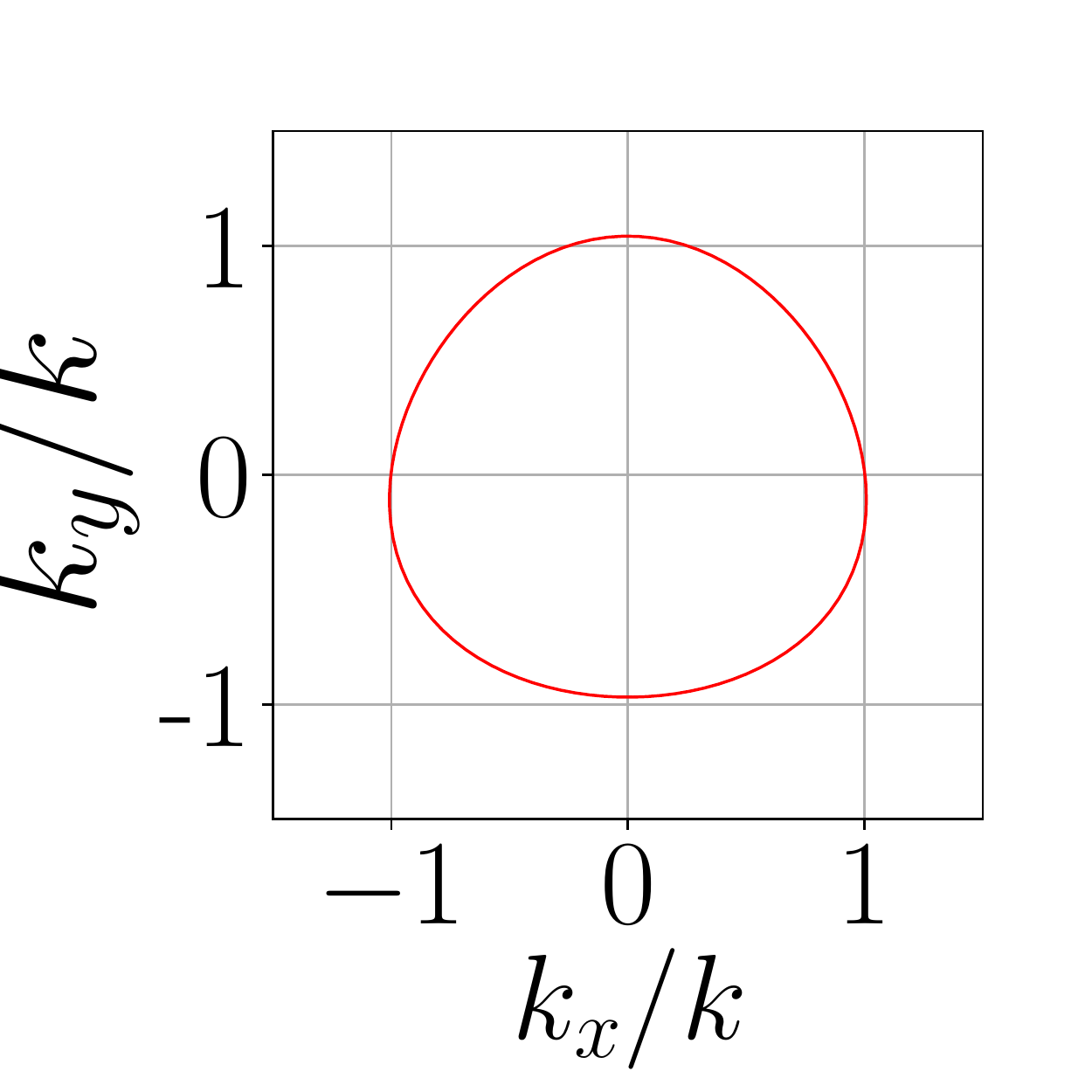}}
     {\includegraphics[width=0.11\textwidth]{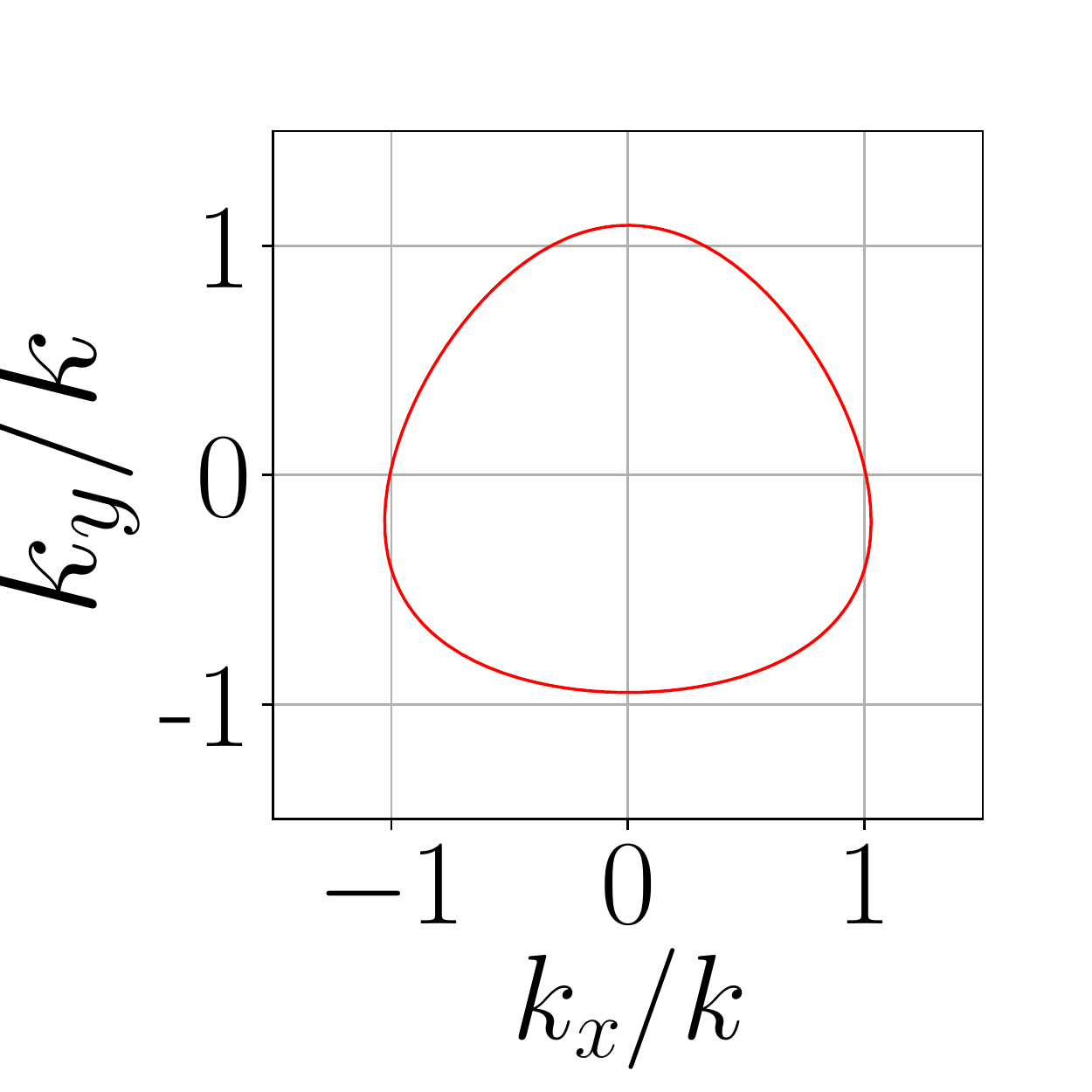}}
     {\includegraphics[width=0.11\textwidth]{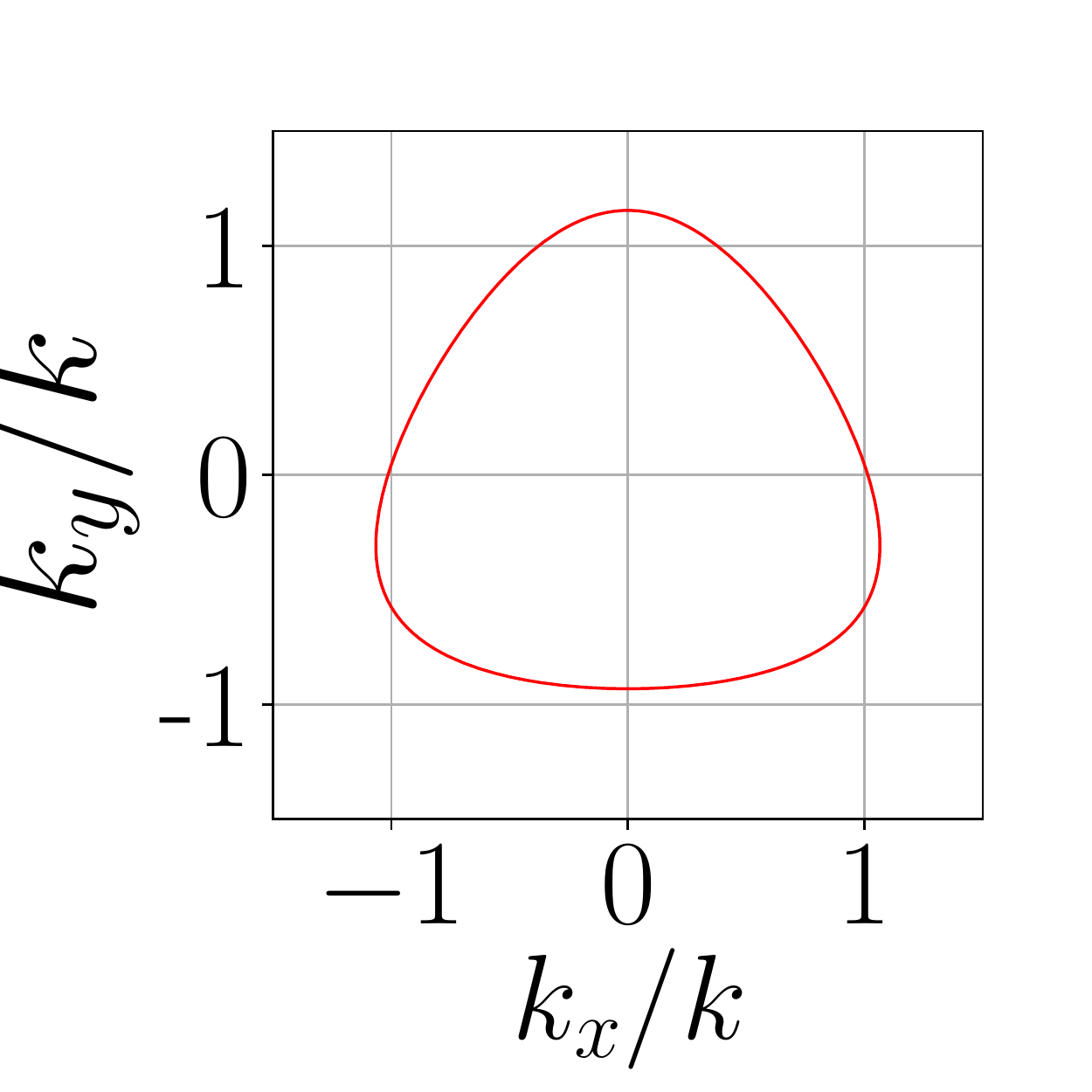}}
     {\includegraphics[width=0.11\textwidth]{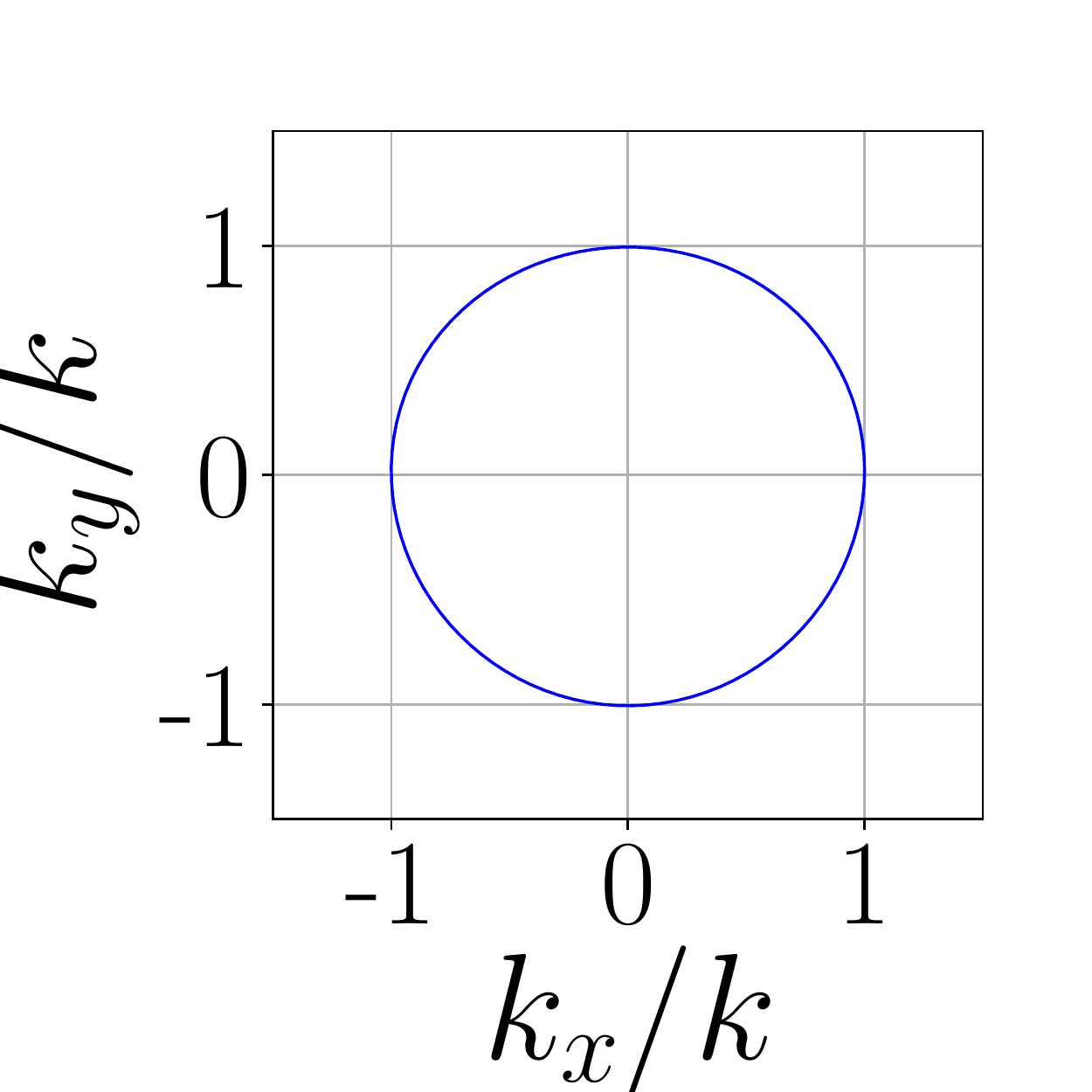}}
     {\includegraphics[width=0.11\textwidth]{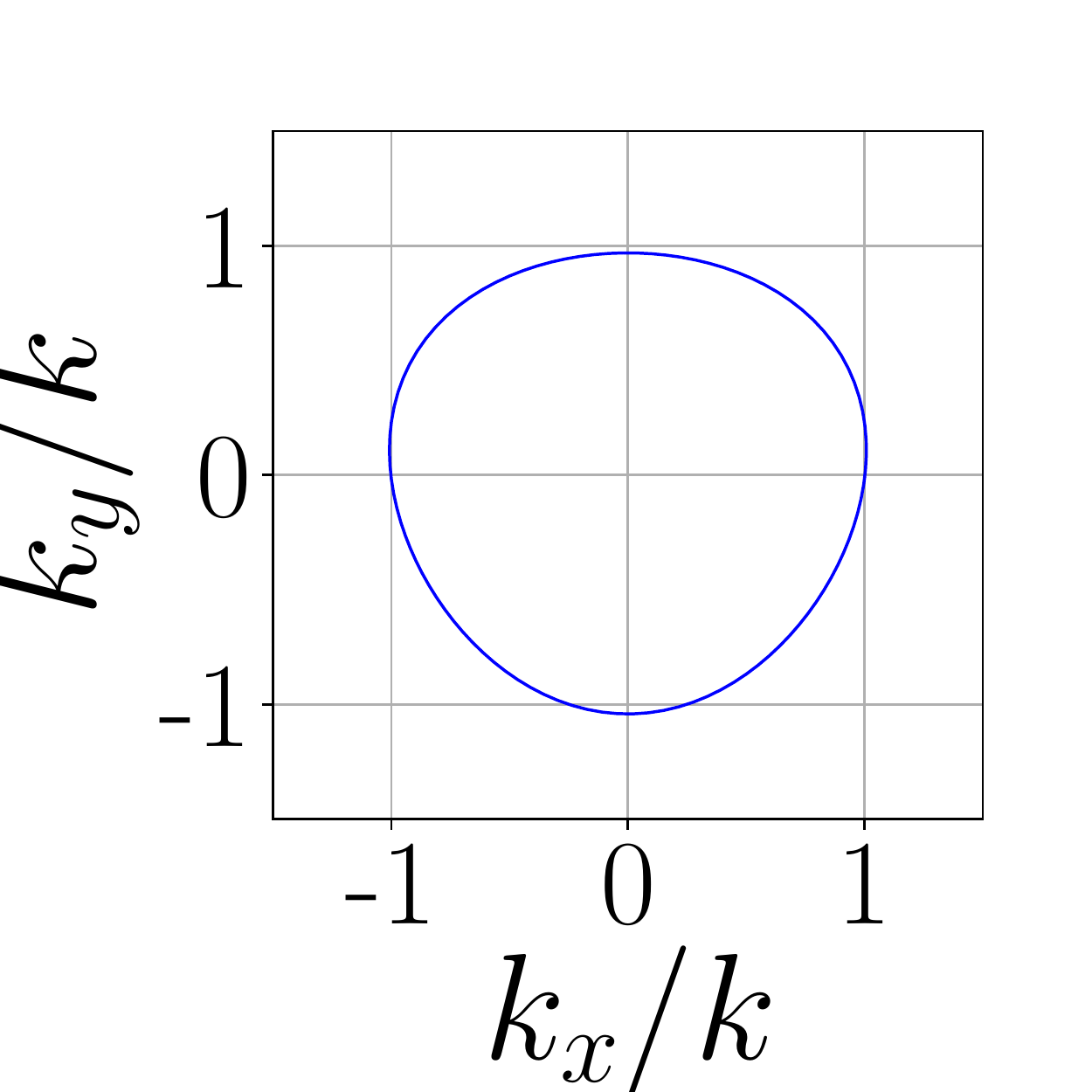}}
     {\includegraphics[width=0.11\textwidth]{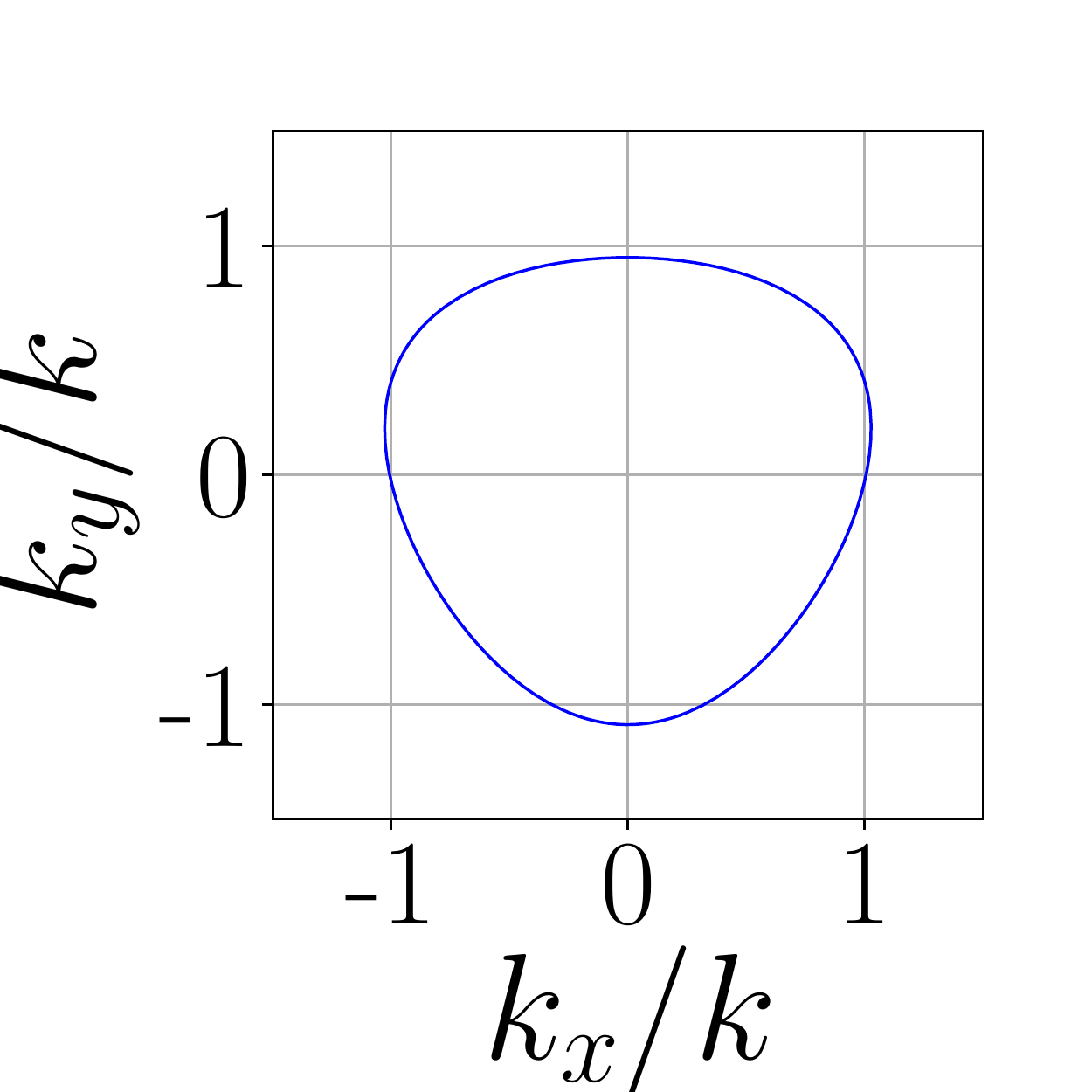}}
     {\includegraphics[width=0.11\textwidth]{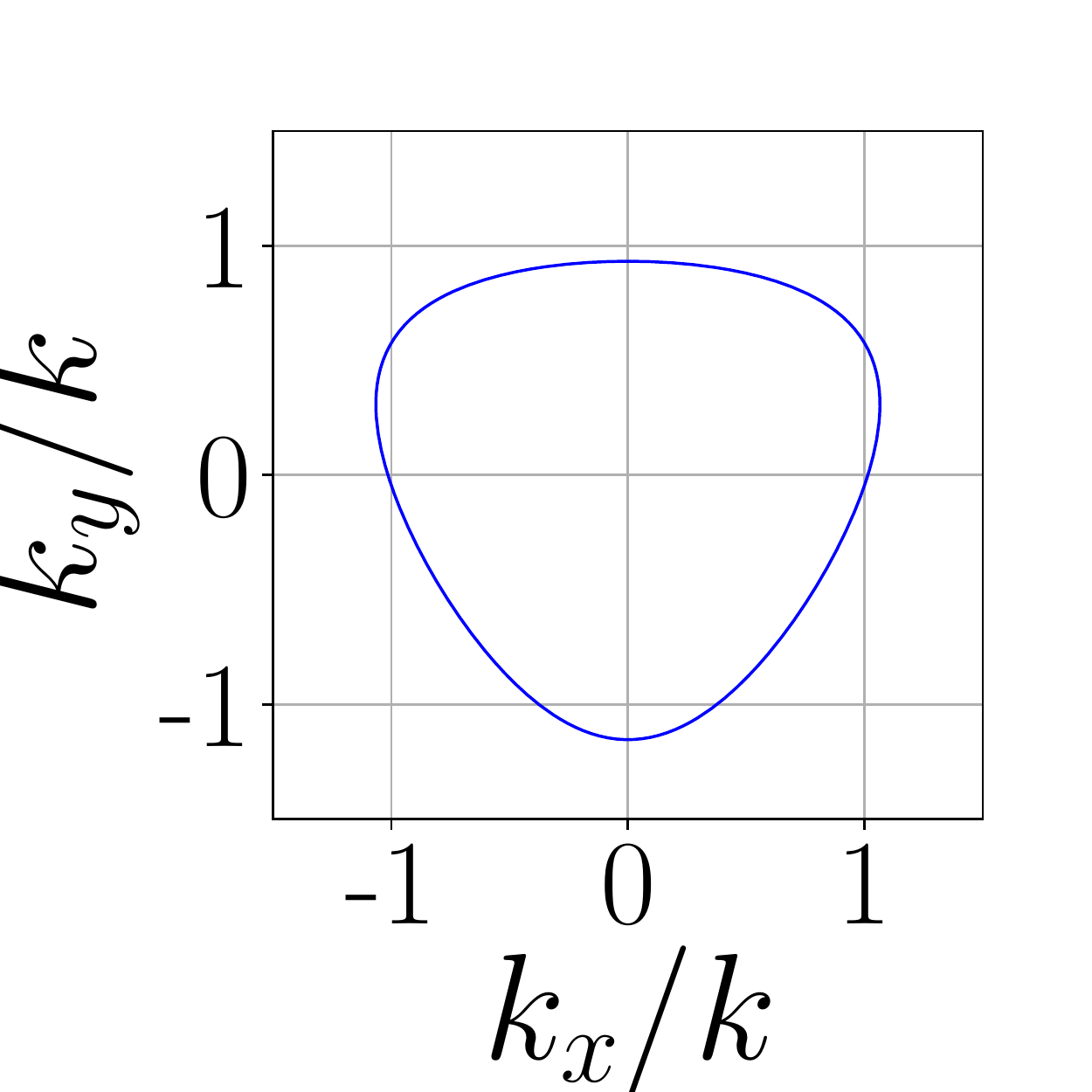}}     
    \caption{The emergence of trigonal warping in the two valleys, with $K$ ($K'$) indicated in red (blue).}
  \label{trigonal}    
  \end{center}
\end{figure}

{\it Valley separation}: The effective Hamiltonian for graphene near the 
charge neutrality point is Dirac-like, ${\cal H} \sim {\bm k} \cdot \boldsymbol{\sigma}$, 
where $\boldsymbol{\sigma}$ are the usual Pauli matrices and reflect the 
two constituent sub-lattices. At low densities, there is a four-fold degeneracy, due to 
spin and valley degrees of freedom. The two inequivalent valleys are located at 
  $K$ and $K'$ respectively in the Brillouin zone, and close to the charge neutrality point
are cylindrically symmetric. For higher densities, the Fermi surface in each valley $K$ and $K'$
exhibits trigonal warping, the emergence of which is shown in Fig. \ref{trigonal}. 


With an applied transverse magnetic field, ${\bm p } \rightarrow \boldsymbol{\pi} = \bm p + e \bm A$, where
$\bm A$ is the vector potential. If the applied field is weak, and the electron or hole density is 
high, the charge carrier dynamics can be described semi-classically, starting from the 
Heisenberg equation of motion for the operators, 
\begin{eqnarray}
\dot{\hat{\boldsymbol{\pi}}}= \left[{\cal H}, \hat{\boldsymbol{\pi}}\right] = e B  \hat{\bm v }\times {\bf n}
\label{eqmotion}
\end{eqnarray}
where ${\bf n}$ is the unit vector normal to the graphene plane, and $B$ is the magnitude of 
the applied magnetic field. Note that $\hat{\boldsymbol{\pi}}$ and $\hat{\bm v}$ are operators. 
Eq. \eqref{eqmotion} is general, and holds for a variety of dispersion relations\cite{Bladwell2015}. In
the semiclassical limit, the operator equation, Eq. \eqref{eqmotion}, is converted to a classical 
equation of the expectation values, which can then be trivially integrated to yield the real space
motion of a electron under an applied transverse magnetic field, 
\begin{eqnarray}
{\bm r}(t) = \frac{\boldsymbol{\pi} \times \bm n}{eB} 
\label{eqmotion1}
\end{eqnarray}
where $\bm r = \left< {\hat{ \bm r}}\right>$ and $\boldsymbol{\pi} = \left<\hat{\boldsymbol{\pi}}\right> $. 
This is the equation of motion for cyclotron motion, with the electron following the equienergetiuc contours 
of the Fermi surface. Thus at high densities, the semi-classical cyclotron orbits of graphene are trigonally warped. 

\begin{figure}[t!]
    \begin{center}
     {\includegraphics[width=0.22\textwidth]{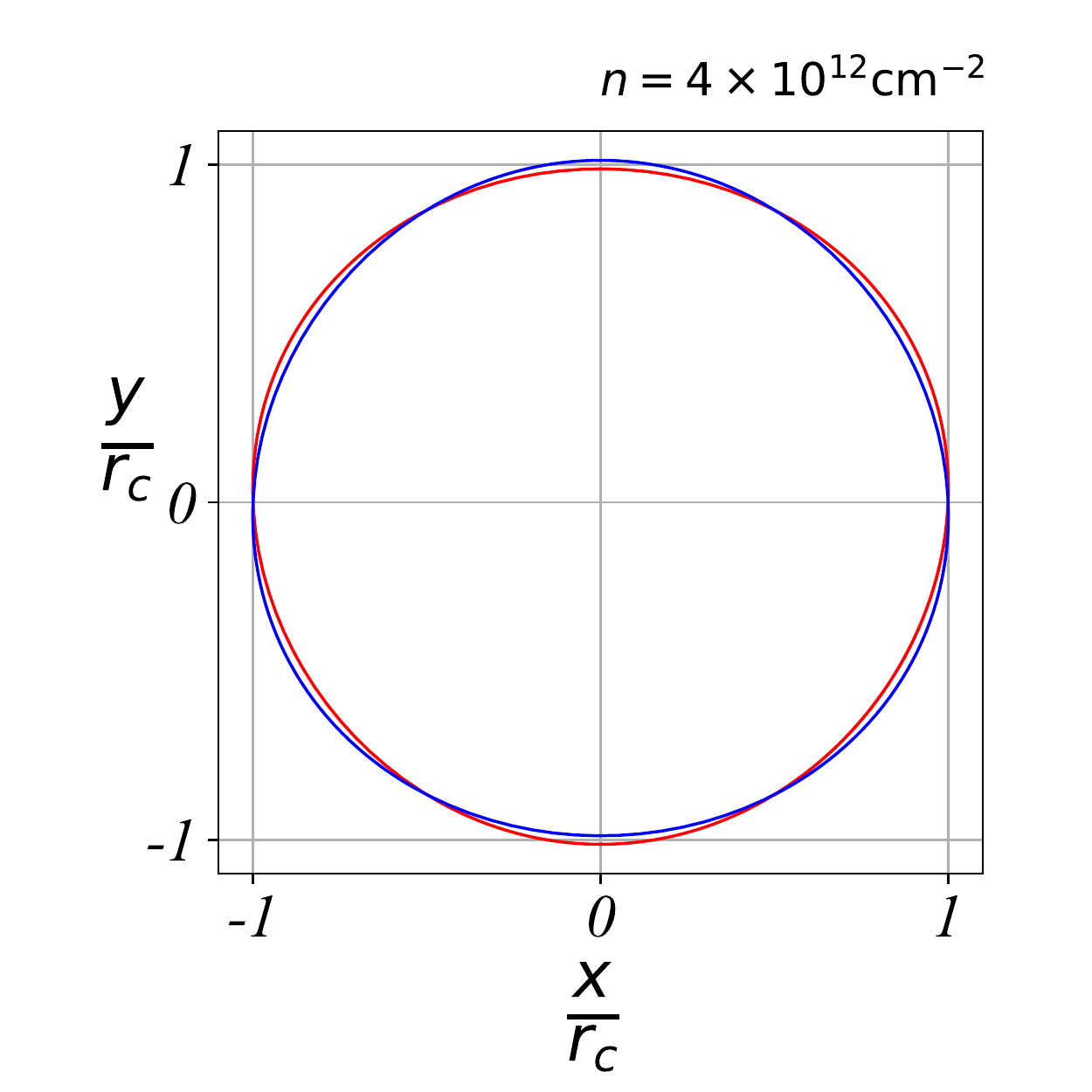}}
     {\includegraphics[width=0.22\textwidth]{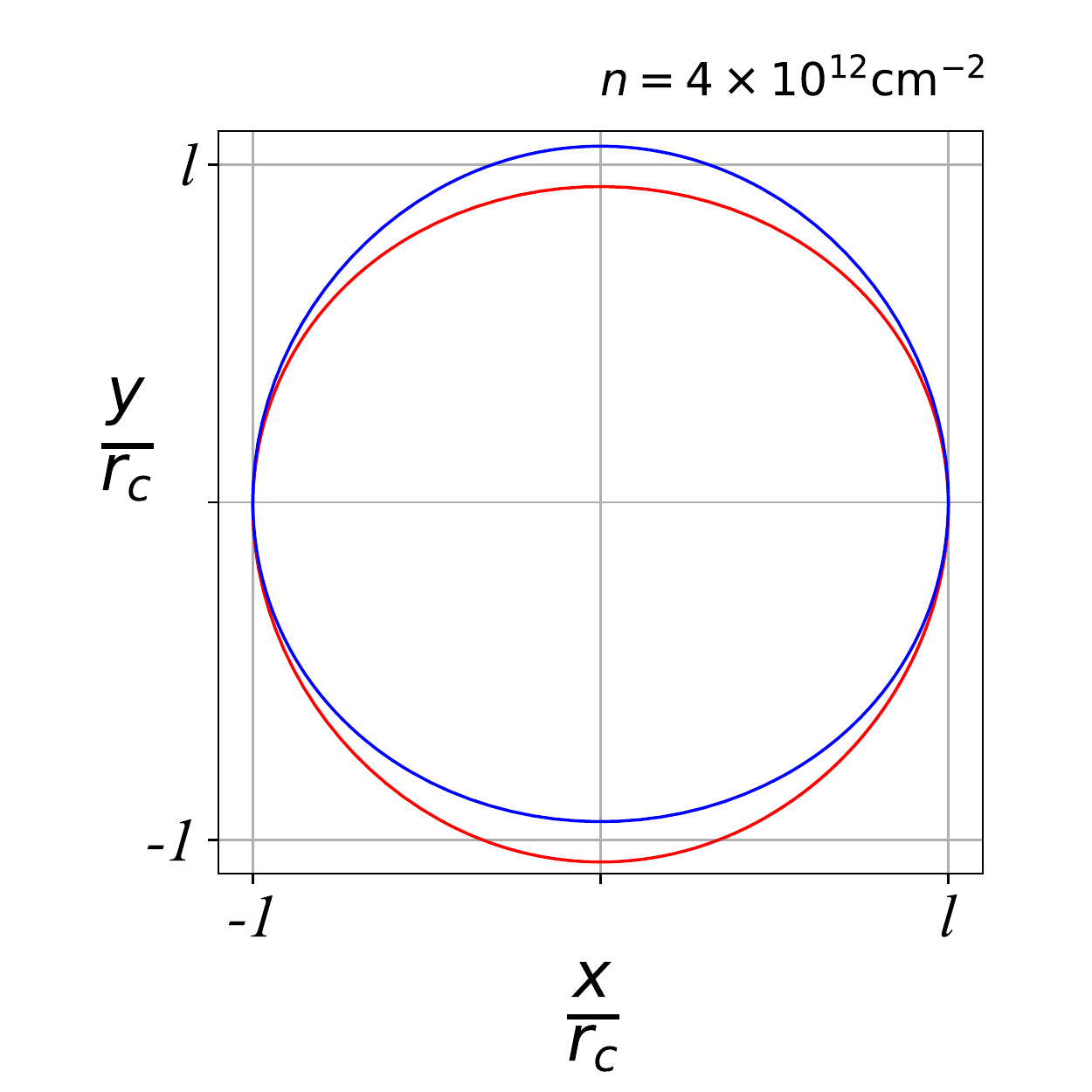}} 
    \caption{
     Left panel: Real space trajectories without the spatial offset induced due to collimation. The spatial separation between the two valleys is 
     small over the entire trajectory. 
    \newline
    Right panel: Real space trajectories including the induced spacial offset from collimation, Eq. \eqref{offset}. The strong enhancement in the 
    spatial offset between the two valley states make resolution of the individual valleys possible. 
    }
  \label{fig2}    
  \end{center}
\end{figure}

In Eq. \eqref{eqmotion1}, guiding centres of the two valleys are identically located at $(x, y) = (0, 0)$. This 
is shown in Fig. \ref{fig2}. 
For an electron optics device, for example, the pin-hole collimator designed in Ref. \cite{Barnard2017}, 
the initial position of the wave packet is at $(0,0)$, the location of the 
injector. In addition, for a perfectly collimated beam of electrons, the initial velocity is fully aligned along
the $x$ axis, parallel to the channel of the injector. For a cylindrically symmetric Fermi surface, the 
location of the 
guiding centre co-ordinate is unchanged. When the Fermi contour becomes trigonally
warped, the guiding centre co-ordinate becomes offset, proportional to the magnitude of the 
trigonal warping. This offset effect is presented in Fig. \ref{fig2}. Since the 
two valleys exhibit triagonal warping with opposite signs, the guiding centre co-ordinates are offset
above and below the $y$ axis. The magnitude of the offset is proportional to the magnitude of the 
trigonal warping; as trigonal warping increases, so does guiding centre co-ordinate offset. 

To illustrate the effect analytically, I consider the following approximation for the Fermi momentum, $p_F$, 
\begin{eqnarray}
p_F \approx \hbar k (1  + s u \sin3\theta ) 
\label{momentum}
\end{eqnarray}
where $\theta$ is the polar angle, $\theta = \tan^{-1} k_y/k_x$, and $k_0 = \sqrt{\pi n}$. The valley index, $s=\pm1$, with $u = k a/4$, where $a$ is the lattice constant
of graphene. It is important to 
note that this analytic approach is only valid while $3u \ll 1$, that is, $u \sim 0.1$. 
 The 
transverse velocity must vanish for collimated injection. From Eq. \eqref{eqmotion1}, the condition is
 ${\partial p_y}/{\partial \theta} = 0$, where $p_y = p_F \cos\theta$. The guiding centre
 co-ordinate is 
 \begin{eqnarray}
  (x_{gc}, y_{gc}) \approx \frac{\hbar k_0}{eB} \left(-s 3 u  , 1 - u \right)
 \label{offset}
 \end{eqnarray} 
 where  $s$ is once again the valley index. This offset of the two valley states can be clearly seen in Fig. \ref{fig2}.

\begin{figure}[t!]
    \begin{center}
     {\includegraphics[width=0.22\textwidth]{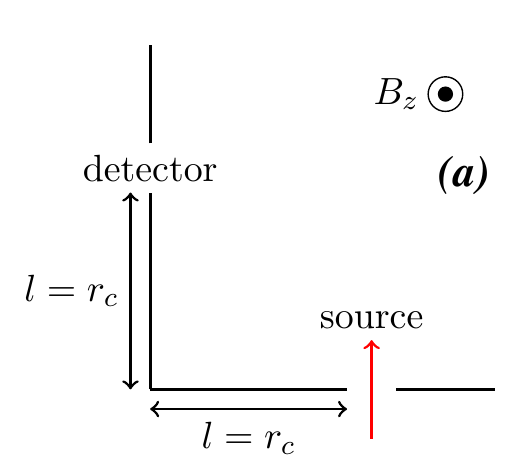}}
     {\includegraphics[width=0.22\textwidth]{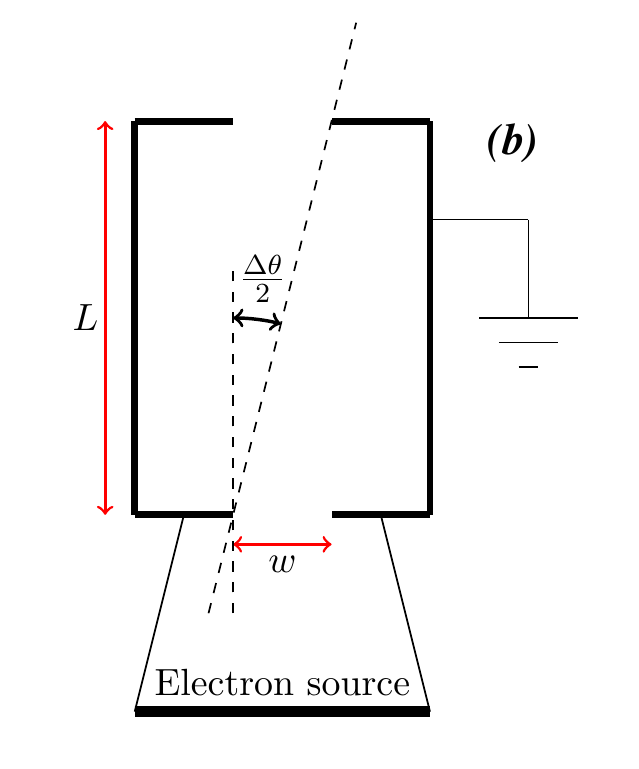}} 
     {\includegraphics[width=0.23\textwidth]{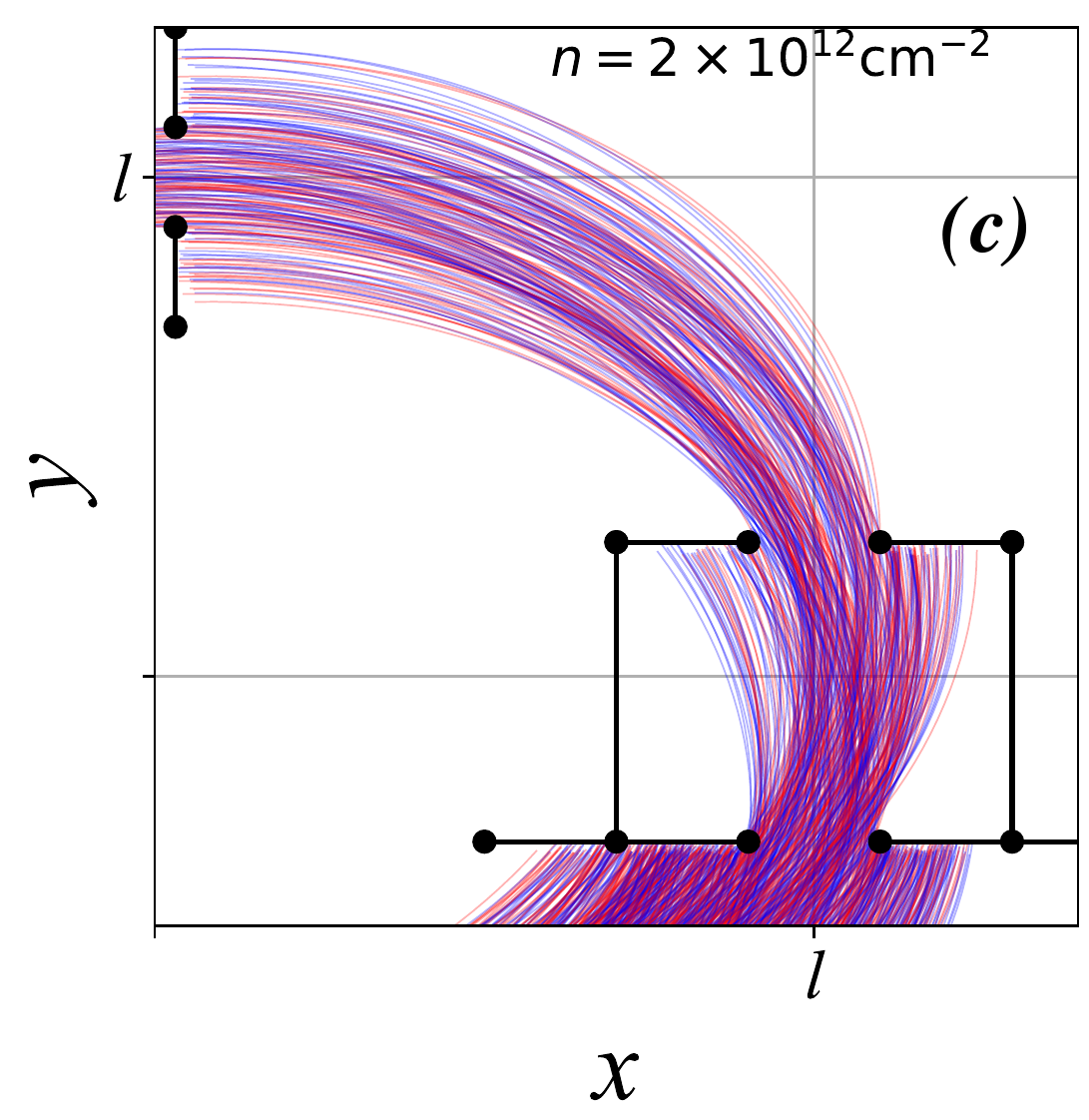}}
     {\includegraphics[width=0.23\textwidth]{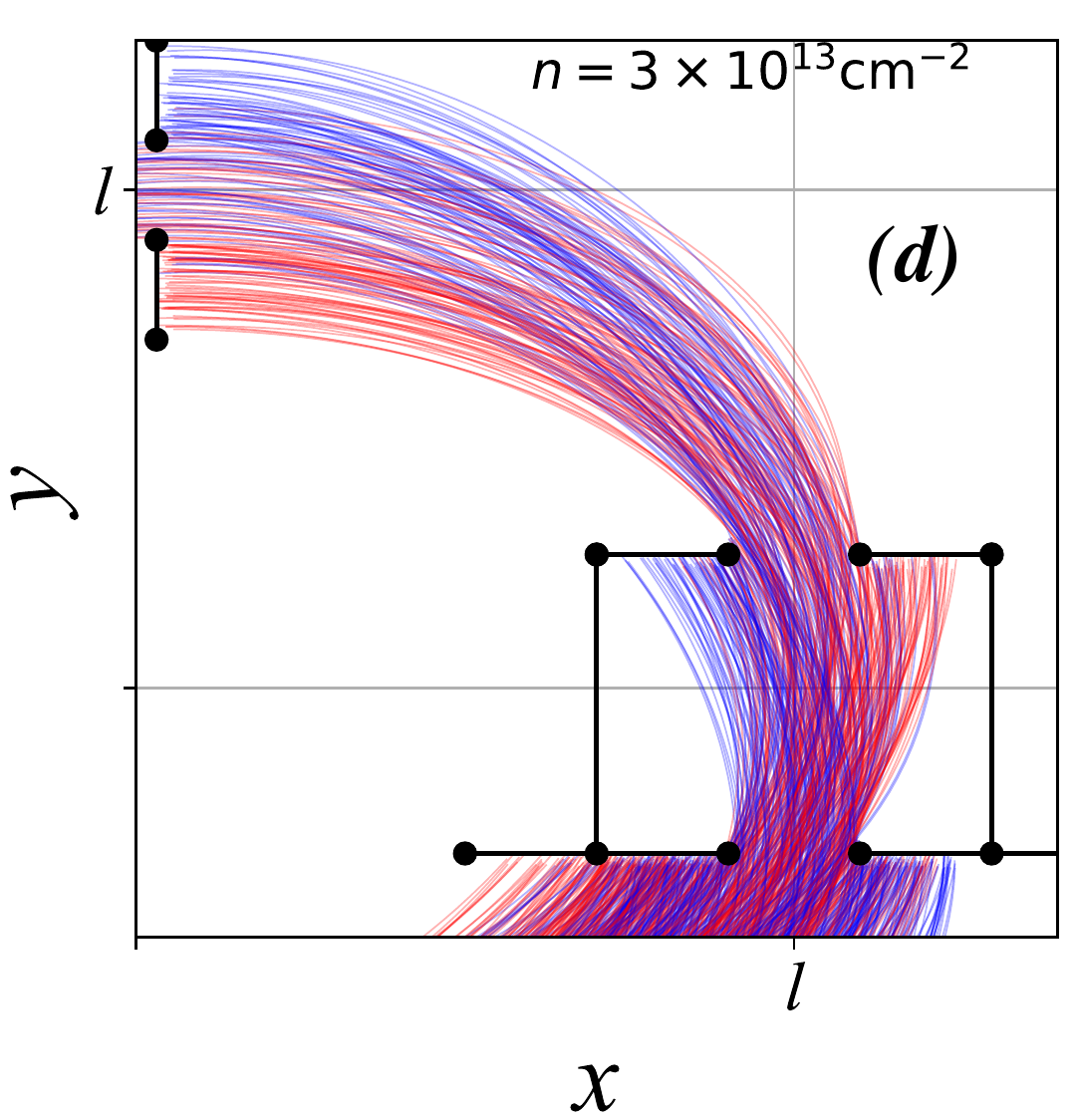}}     
     \caption{Top left panel: Orthogonal focusing setup, with source and detector planes oriented orthogonally. Here $l$ is the focusing length, 
    and $B_z$ the weak focusing field. 
    \newline
    Top right panel: Pinhole collimator layout. The aperture width is $w$, and the length $L$, with the angular spread, $\Delta \theta$ given by the twice the projected 
    angle. This is one possible collimating source in graphene, with current experimental devices producing a beam with a beam spread of $\Delta \theta \approx \ang{20}$. 
    \newline
    Lower panels: Trajectory computation for a pinhole collimator source, and pinhole aperture detector, with $l=1.5\mu$m, $L= 900\mu$m, 
    and $w = 300nm$. Trajectories are computed with a random initial condition (random guiding centre coordinate), with the 
    transverse magnetic field, $B = \hbar k_0/eL$. Each valley has trajectories in red (blue). The parameters for the Pinhole collimator are chosen to 
    be comparable to the device of Barnard {\it et al} \cite{Barnard2017}. At very high density the separation between valleys can be seen. High 
    density is required for the weak collimation of the pinhole collimator.  }
  \label{fig3}    
  \end{center}
\end{figure}

Focusing in the half plane, that is, with source and the detector located on the $y$ axis, does not yield any 
spatial separation between the valley states. However, if an orthogonal device setup is employed, as shown in Fig. \ref{fig3}a,
the states can be spatial separated. The real space separation at the collector is
\begin{eqnarray}
x_+ - x_- \approx 8 u \frac{\hbar k_0}{eB} 
\label{realspaceoffset}
\end{eqnarray}
where $x_+$ and $x_-$ are the real space positions in the plane of the detector. 
Thus trigonal warping of the valleys in graphene induces a real space 
separation when combined with an appropriate device setup. { The critical point to note is the significant enhancement; the required density
for valley separation is reduced by a factor of 16, making it possible to resolve the individual valleys at experimentally achievable 
densities.} This real space separation from collimated injection is 
shown in Fig. \ref{fig3}. 

{\it Resolution limits:} On its own, Eq. \eqref{realspaceoffset} tells us little about whether the spatial splitting can be {\it observed}. If the broadening of the beam 
 exceeds the spatial separation, then the valley states will overlap and resolution of the distinct valleys currents will not be possible. There are 
 three specific sources of broadening that are relevant at cryogenic temperatures
 in high mobility samples; (1)  the imperfect collimation of the source, (2) the spatial resolution 
 of the detector, and (3), the beam broadening due to the medium and scattering.

Firstly, we consider the resolution limits due to the imperfect collimation of the source. For a 
 a pinhole style collimator setup in the orthogonal geometry of Fig. \ref{fig3}a, the allowed trajectories 
 are those with guiding centre co-ordinates, 
 $(x_{gc}, y_{gc} \in {\cal S})$, satisfying the limits, $y = L/2, -L/2$, $l -w/2<x< l + w/2$. 
Here $L$ and $w$ represent the width and length of the pinhole collimator respectively, while $l$ is the focusing 
length. A schematic showing these is presented in Fig. \ref{fig3}b. Specifically, the value of interest is the 
range of $y_{gc}$, which must be less than the spatial separation of the valley states. Provided the trigonal warping is relatively small, $u < 0.1$, 
the radius of curvature in the local region about $\left<v_x\right> = 0$ is identical for both valleys, and the trajectories within
the collimator can be approximated by cyclotron orbits. The resulting approximate limit is $|y_{gc}| < \sqrt{wr_c} - L/2$, while the 
``spread" of the beam is double this. 
In the absence of any addition collimation, 
the required density for spatial separation for a pinhole collimator 
directly equivalent to that of Barnard {\it et al} is $w = r_c/6$ and $L = r_c/2$ is $n \approx 4\times 10^{13}$cm$^{-2}$. 
Alternatively, the spread can be determined from typical values in the literature. For a pinhole collimator like that of Barnard {\it et al}, 
the angular full width half maximum was $\Delta \theta \sim \pi/9$, corresponding to a spread of $\Delta x = \Delta \theta r_c$. For 
$x_+ - x_- > \Delta x$, $8u > \pi/9$, which yields a minimum density of $n \approx 3 \times 10^{13}$cm$^{-2}$. 
{  In general, the angular FWHM, $\Delta\theta$ scales as $w/L$, thus the
required density for the resolution of individual valleys scales as 
\begin{eqnarray}
n \approx \left(\frac{w}{L}\right)^2 4 \times 10^{-18}{\text{cm}}^{-2}
\label{res1}
\end{eqnarray}
Lithographic limits for feature sizes are typically $\sim 100$nm. 
}

{ Parabolic {\it p-n} junctions, 
as proposed by Liu {\it et al} have a significantly more collimated beam shape, with $\Delta \theta \sim \ang{5} \approx 1/10$. Implementation using this 
type of beam gives $8 u > 1/10$, implying an electron density for resolution of the valleys of
 $n \sim 4  \times 10^{12}$cm$^{-2}$, well within the bounds of back-gated hexagonal 
Boron-Nitride devices. 
The limitation of this form of collimation is the larger beam expanse, $\Delta y \sim 400nm$. The corresponding spread of the beam
at the detector is $\Delta x \approx (\Delta y/l)^2/2$, which for focusing lengths of $l = 1\mu$m is $\Delta x \sim 1/8$. Due to the 
quadratic scaling, the beam expanse becomes unimportant for larger $l$. 
Combinations of parabolic junctions and 
pinhole aperture can further improve the angular distribution and spatial extent of the beam\cite{Boggild2017}. 
}

{  Next, we consider the resolution of the detector. By far the simplest setup is a pinhole aperture, with the voltage probe connected behind. 
For resolution of the individual valleys, the requirement is that the spatial separation of the 
valley peaks, $8u r_c$, is greater than the width of the pinhole aperture, $w$. For grounded contacts, the limitation on $w$ is lithography, 
$w \sim 100$nm. For a focusing length $l \sim 1\mu$m, the ratio of the pinhole aperture to the focusing length can be $w/l \sim 1/10$. 
If both $\Delta \theta \sim 1/10$ and $w/l\sim 1/10$, the required density for the resolution of individual valleys is
$n \sim 4 \times 10^{12}$cm$^{-2}$, remarkably close to the densities of Barnard {\it et al}. 

For such a device, a final question remains as 
to whether beam broadening due to the medium destroy resolution of the valley peaks. 
The path length of the ballistic measurements used in Ref. \cite{Barnard2017} were
$l_{path} \sim 3-4\mu$m, with negligible beam broadening. 
For an orthogonal device with a geometry equivalent to that of Fig. \ref{fig3}a, this corresponds to a 
maximum focusing length, $l \sim 2\mu$m. Beam broadening due to scattering is therefore unimportant 
at cryogenic temperatures for $l< 2\mu$m. In principle the device can be made significantly larger, 
the mean free path in h-BN encapsulated graphene can be upwards of 
$10\mu$m at densities $\sim 2\times 10^{12}$cm$^{-2}$\cite{Banszerus2016}. 
}

{\it Numerical simulations:} This can be grounded more firmly via numerical simulation of the device, 
 to determine the required $u$ and therefore $n$ for resolution of the individual valleys. 
Since Eqs. \eqref{realspaceoffset} and \eqref{momentum} are valid only for small values of trigonal warping, 
I consider the energy bands of the usual tight-binding hamiltonian (see, for example, \cite{Neto2009}) to 
determine the equienergetic contours and then use Eq. \eqref{eqmotion1} to determine the dynamics.
As already noted, this approach is valid provided $k r_c > 100$, and device feature
sizes much large than the Fermi wavelength. We will consider both a pinhole collimator, equivalent to that
shown in Fig. \ref{fig3}, and a parabolic {\it p-n} junction\footnote{See supplementary material {  [at url] } for the python code used to generate the figures. Code 
comments provide details of the exact procedure.}.

 \begin{figure}[t!]
    \begin{center}
     {\includegraphics[width=0.22\textwidth]{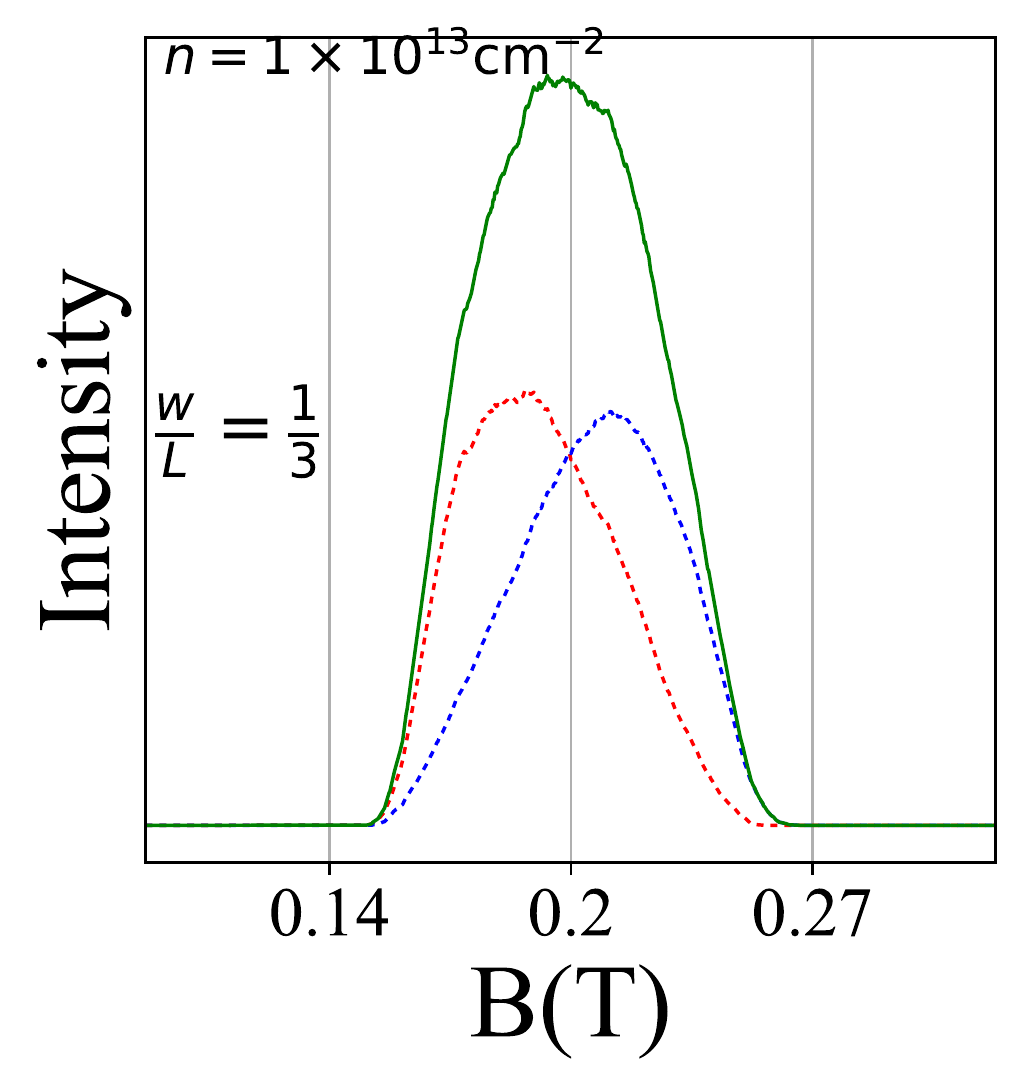}}
     {\includegraphics[width=0.22\textwidth]{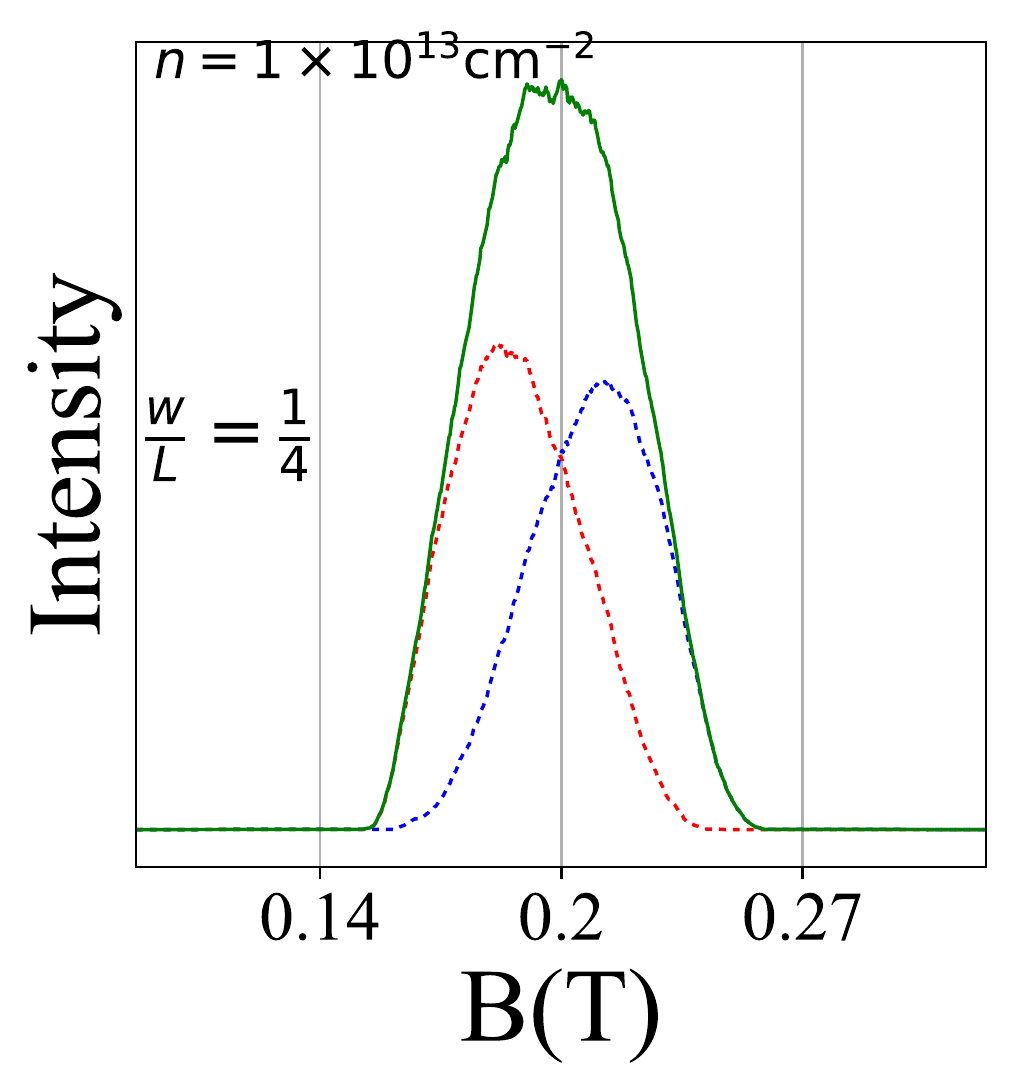}} 
     {\includegraphics[width=0.22\textwidth]{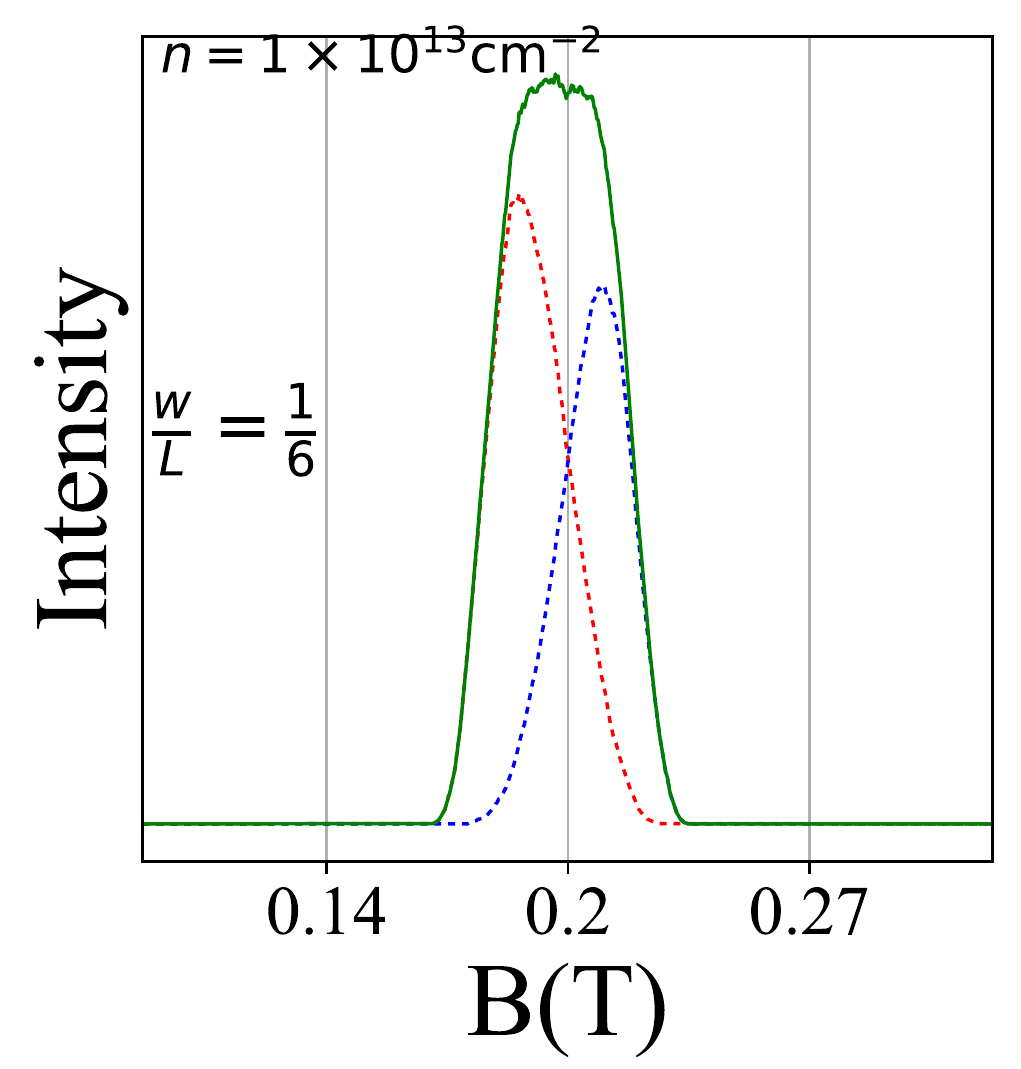}}
     {\includegraphics[width=0.22\textwidth]{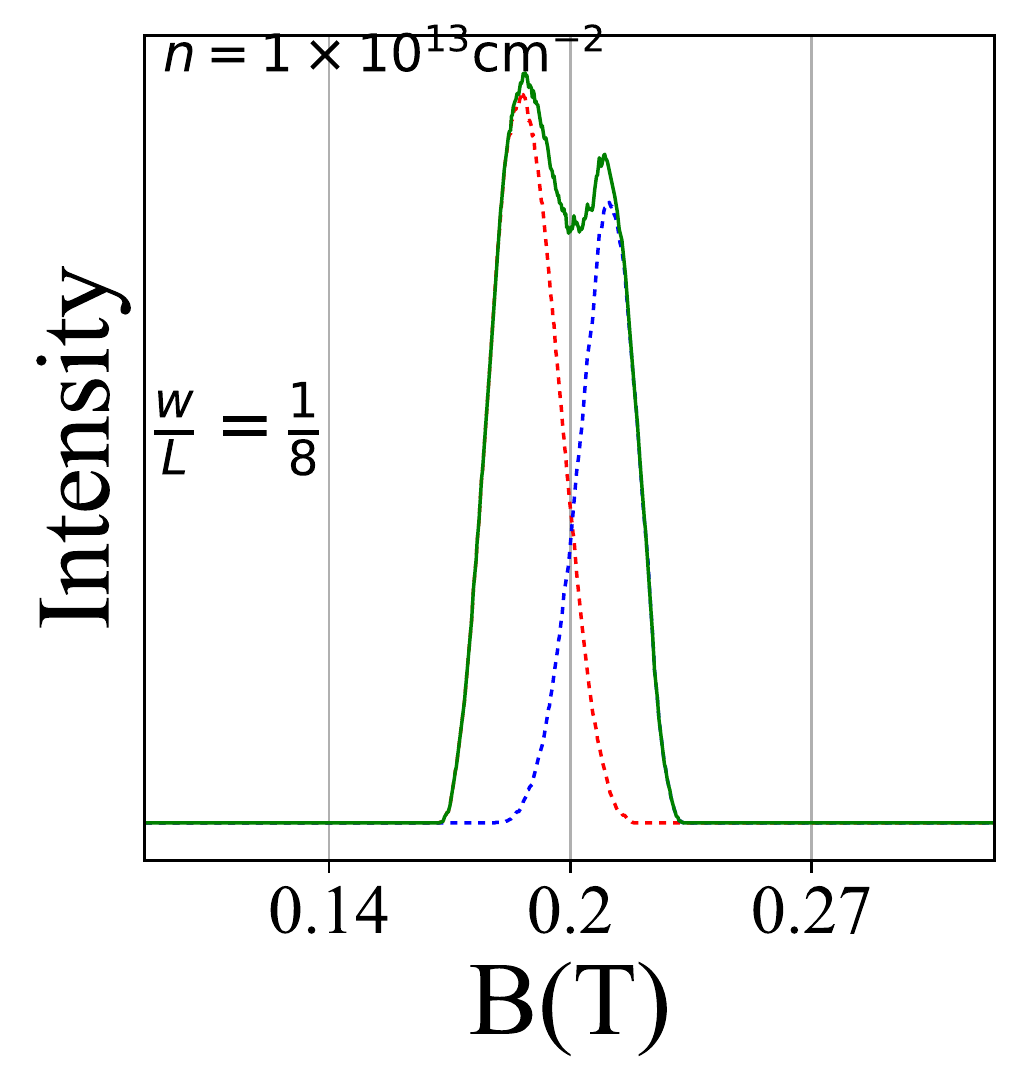}}     
    \caption{Results of numerical simulations of a device with $l = 1.5\mu$m, with varying resolution factor $\rho = w/L$. 
    The injector consists of a pinhole collimator, with the detector a pinhole aperture. Densities are chosen as the lowest 
    allowing resolution of the peaks.
    The density is fixed at $n = 1 \times 10^{13}$cm$^{-2}$.  
   The results for the resolution of individual valley peaks agree with Eq. \eqref{res1}.
   Here the red and blue dashed curves correspond to the signal from each valley, 
    with the green curve being the total signal. }
  \label{numerics}    
  \end{center}
\end{figure} 

The results of the numerical procedure are presented in Figs. \ref{numerics}. For a device with a pinhole collimator, 
and a detector with a simple pinhole aperture equivalent to that of 
Barnard {\it et al}, the required density is greater than $3 \times 10^{13}$cm$^{-2}$, as would be anticipated based
on Eq. \eqref{res1}. With narrower
apertures lowering the required density. At $w/L = 1/8$, the peaks can be clearly resolved at $n = 1 \times 10^{13}$cm$^{-2}$. 
This is well within current experimentally achievable 
densities for graphene devices\cite{Craciun2011}, however it requires specific gating techniques. 
Further improvements by reducing $w$ are possible. In principle 
there is a fundamental limit due to diffraction at low densities, however for $n =  1 \times 10^{13}$cm$^{-2}$, $\lambda_F \sim 10$nm $\ll w = 100$nm, 
and diffraction is essentially irrelevant. 

\begin{figure}[t!]
    \begin{center}
     {\includegraphics[width=0.22\textwidth]{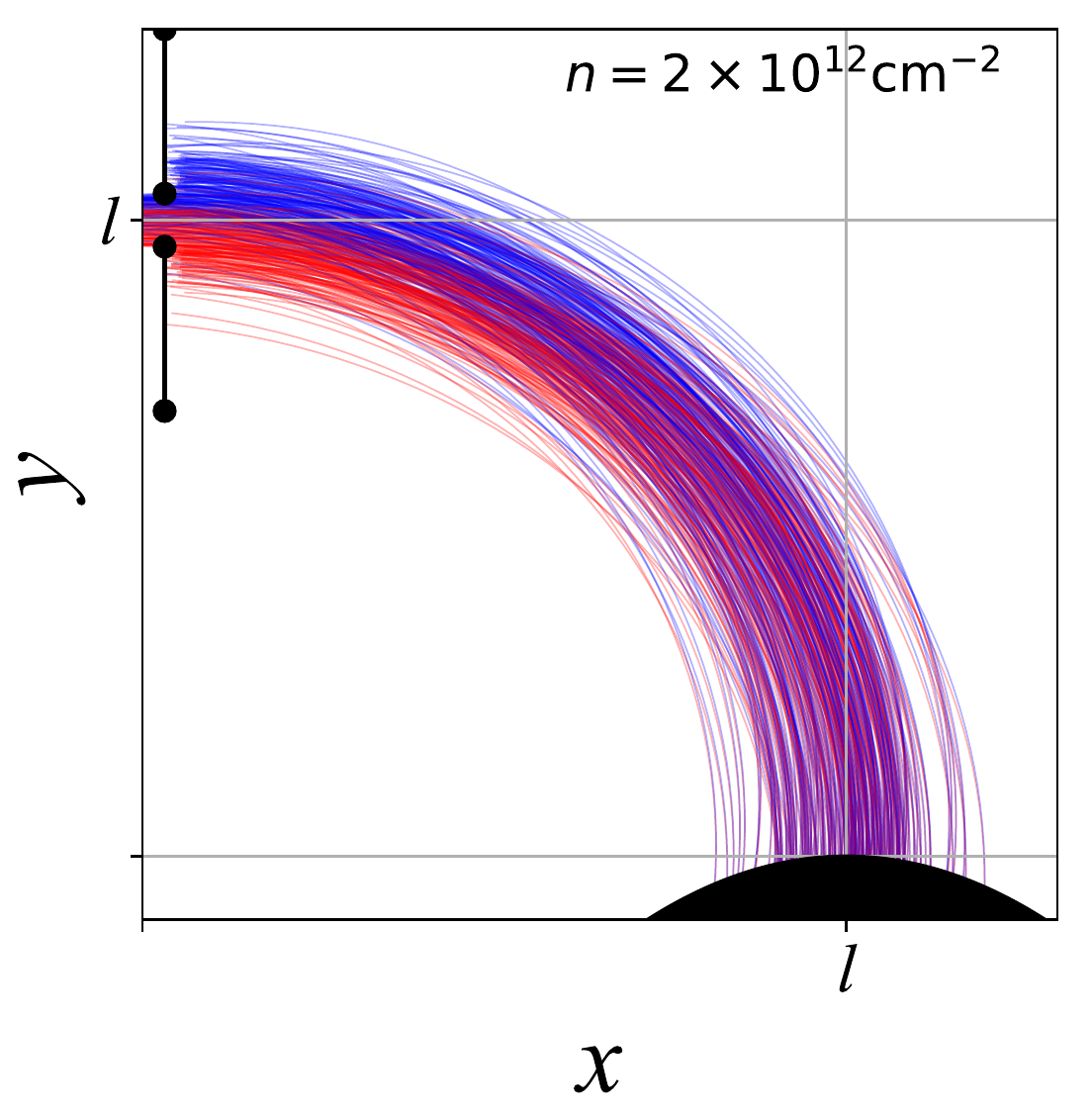}}
     {\includegraphics[width=0.22\textwidth]{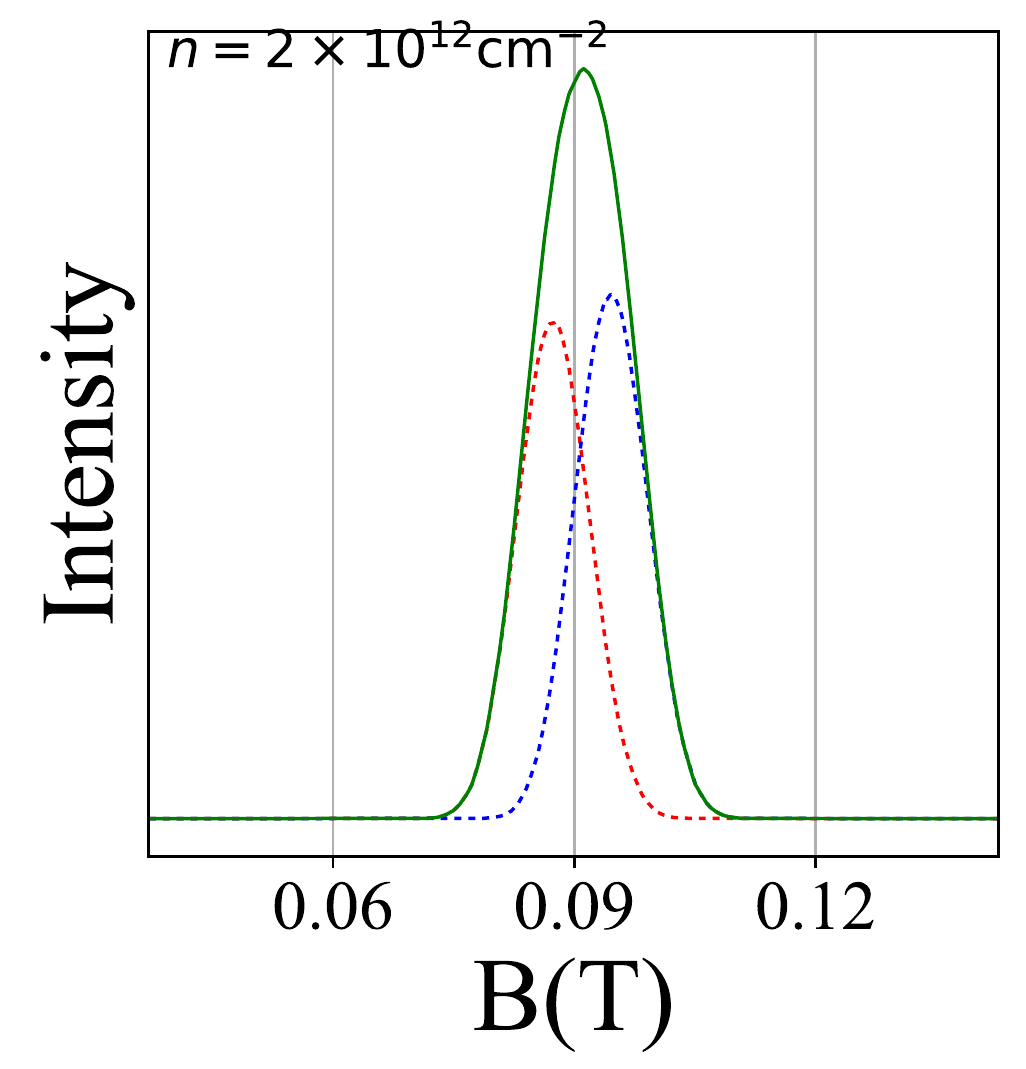}} 
     {\includegraphics[width=0.22\textwidth]{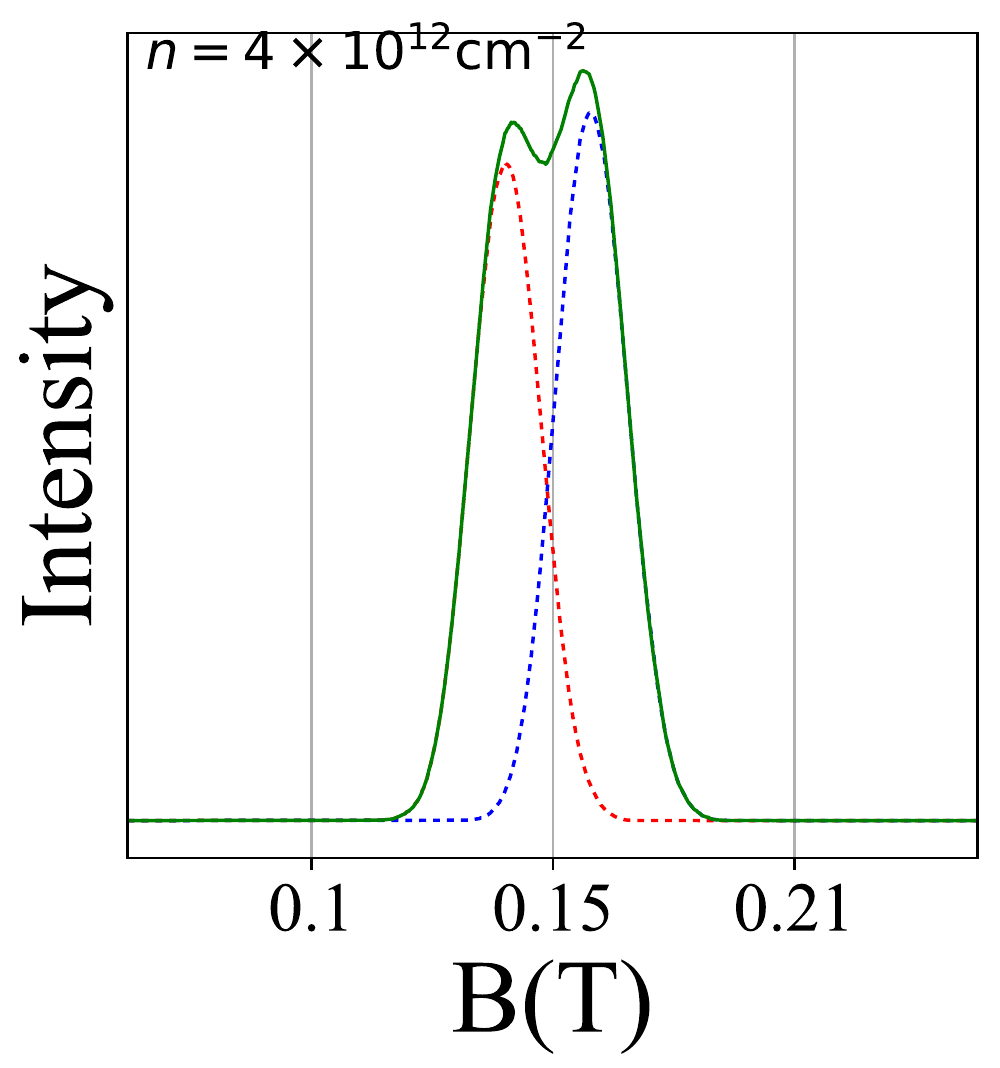}}
     {\includegraphics[width=0.22\textwidth]{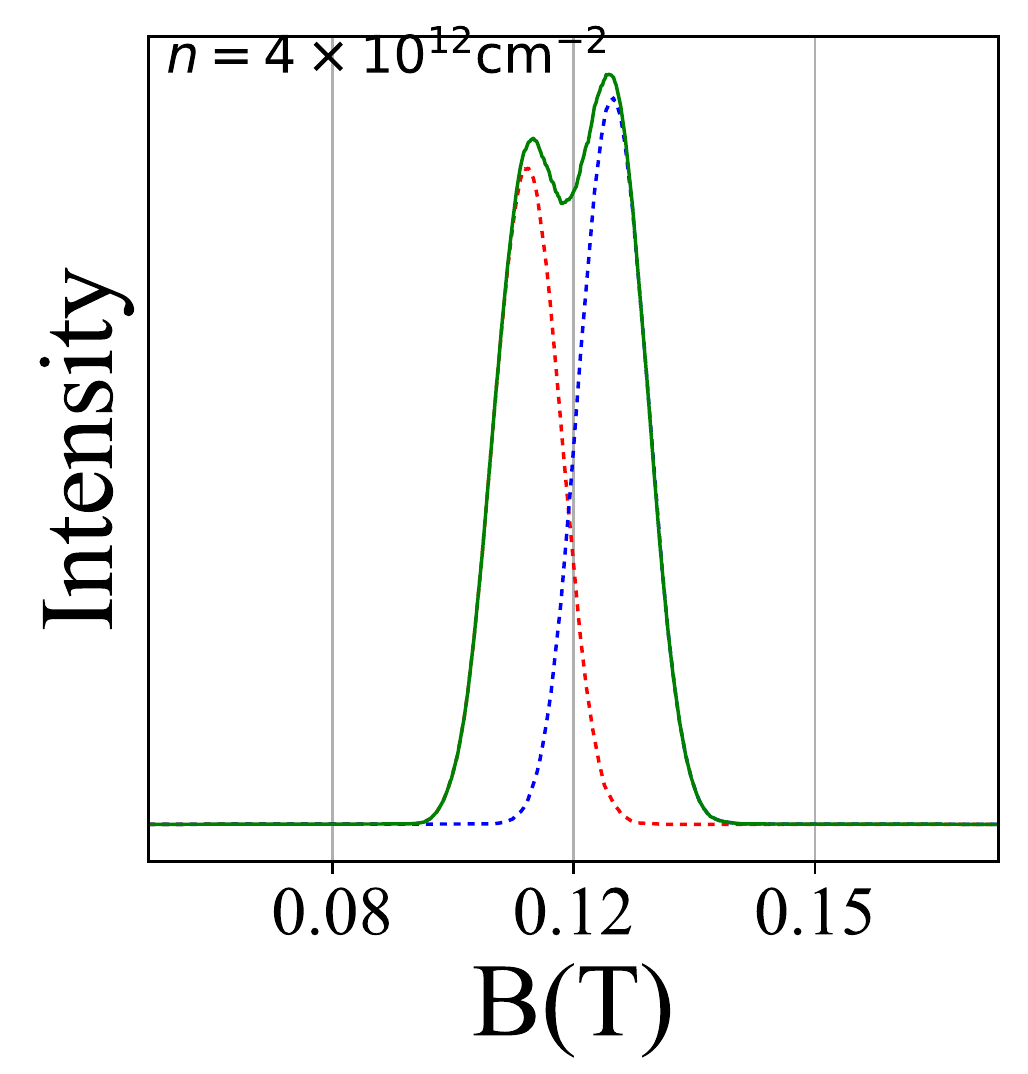}}     
    \caption{
    Top left panel: Trajectories from numerical simulations of a device with $l = 1.5\mu$m, a parabolic PN junction with a focal length of 
    400nm, and pinhole style detector of width $w = 150$nm. Red and blue traces indicate different valleys. The parabolic PN junction
    source is shaded in black. 
    \newline
    Top right panel: Focusing spectrum for the parabolic PN junction, with $l = 1.5\mu$m, $w = 150$nm, $n = 2 \times 10^{12}$cm$^{-2}$. 
    \newline
    Lower left panel: Focusing spectrum for the parabolic PN junction, with $l = 1.5\mu$m, $w = 150$nm, $n = 4 \times 10^{12}$cm$^{-2}$. 
    \newline
    Lower right panel: Focusing spectrum for the parabolic PN junction, with $l = 2\mu$m, $w = 150$nm, $n = 4 \times 10^{12}$cm$^{-2}$. Increasing
    the focusing length increases the separation of the valley states, and thereby improves resolution. 
     }
  \label{numerics2}    
  \end{center}
\end{figure}

{  Next, we consider a parabolic PN junction for current injection, with the pinhole aperture for detection of the current. The 
considered parabolic PN junction is equivalent to that of Liu {\it et al}, with a gaussian beam width of $400$nm, and 
a full width half maximum angular spread of $\ang{5}$\cite{Liu2017}.  We consider a 
 pinhole aperture detector of width $w = 150$nm. The results of the numerical simulation are presented in
 Fig. \ref{numerics2}. The valley split peaks can be resolved at $n = 4 \times 10^{12}$cm$^{-2}$, which
 is accessible in high mobility back gated devices. With an increased the focusing length, the resolution improves, due to the 
 increase in the real space separation of the valley states. Finally, we note that the relative valley warping at $n = 4 \times 10^{12}$cm$^{-2}$
 is extremely small, $u \approx 0.013$. That individual valley states can be resolved is an illustration of the significant enhancement 
 present in Eq. \eqref{realspaceoffset}. It is this enhancement that makes the valley separation possible. 
 }

In conclusion, we have shown that the ballistic trajectories of electrons in different valleys of graphene can be spatial 
separated using a collimating source, and a weak magnetic field. With an appropriately chosen geometry, the offset in 
different valleys due to trigonal warping is strongly enhanced. 
The magnitude of the valley separation at $\sim 4 \times 10^{12}$cm$^{-2}$
is sufficient to fully resolve the individual peaks when the source is a parabolic PN junction, 
and this carrier density is achievable with current gating methods, in high mobility back gated devices.
For pinhole collimators, resolution is lower, due to the greater spread of the beam, and the density requirements for 
experimentally feasible devices are therefore larger.
Finally, the basic principle 
of valley separation outlined here will work for any two dimensional systems where the valleys exhibit trigonal warping, including
bilayer graphene, moire superlattices, and two dimensional transitional metal dichalcogenides, and all will exhibit this enhancement. 
This research was partially supported by the 
Australian Research Council Centre of Excellence in Future Low-Energy Electronics Technologies (project number CE170100039) 
and funded by the Australian Government. SSRB would like to thank Oleg Sushkov for his critical reading and suggestions.

\bibliography{Biblio.bib}

\end{document}